\documentclass[a4paper]{article}
\usepackage[psamsfonts]{amssymb}
\usepackage{amsmath}
\usepackage{latexsym}
\usepackage{pstricks}
\usepackage{cite}
\usepackage{epsfig}
\date{}

\author{M. Alimohammadi\footnote{alimohmd@ut.ac.ir}\ \ and N.
Olanj
\\ {\small Department of Physics, University of Tehran,}
\\ {\small North Karegar Ave., Tehran, Iran.}}
\title{Class of solvable reaction-diffusion processes on Cayley tree }
\begin{document}
\maketitle
\begin{abstract}
Considering the most general one-species reaction-diffusion
processes on a Cayley tree, it has been shown that there exist two
integrable models. In the first model, the reactions are the
various creation processes, i.e.
$\circ\circ\rightarrow\bullet\circ$,
$\circ\circ\rightarrow\bullet\bullet$ and
$\circ\bullet\rightarrow\bullet\bullet$, and in the second model,
only the diffusion process $\bullet\circ\rightarrow\circ\bullet$
exists. For the first model, the probabilities $P_l(m;t)$, of
finding $m$ particles on $l$-th shell of Cayley tree, have been
found exactly, and for the second model, the functions $P_l(1;t)$
have been calculated. It has been shown that these are the only
integrable models, if one restricts himself to $L+1$-shell
probabilities $P(m_0,m_1,\cdots,m_L;t)$s.
\end{abstract}
\section{Introduction}
The integrable reaction-diffusion processes have been investigated
by various methods on one-dimensional lattice. A class of these
models are characterized by a master equation with appropriate
boundary conditions. These boundary terms, which determine the
probabilities at the boundary of the space of the parameters, are
chosen such that the studying of the various reactions becomes
possible via a simple master equation. The basic quantity in these
models is the conditional probability
$P(\alpha_{1},\cdots,\alpha_{N},x_{1},\cdots,x_{N};t|\beta_{1},
\cdots,\beta_{N},y_{1},\cdots,y_{N};0)$, which is the probability
of finding particles $\alpha_{1},\cdots,\alpha_{N}$ at time $t$ at
sites $x_{1},\cdots,x_{N}$, respectively, if at $t=0$ we have
particles $\beta_{1},\cdots,\beta_{N}$ at sites
$y_{1},\cdots,y_{N}$, respectively \cite{1,2,3,4,5,6}.

Another class of models are those which are solvable through the
empty interval method. In these models, the main quantity, in the
most general case, is $E_{k,n}(t)$ which is the probability that
$n$ consecutive sites, starting from the site $k$, are empty.
Several examples have been investigated by this method
\cite{7,8,9,10,11,12,13}.

The crucial property which causes the above mentioned cases to be
analytically solvable, is the dimension of the lattice which
particles move on it. One dimensional lattice is clearly the
simplest case.

Considering the higher-dimensional lattice in studying the
integrable models is obviously a great improvement in theoretical
physics. One of the important example in this area is Cayley tree.
A Cayley tree is a connected cycle-free graph where each site is
connected to $z$ neighbor sites. $z$ is called the coordination
number. The Cayley tree of coordination number $z$ may be
constructed by starting from a single central node (called the
root of the lattice) at shell $l=0$, which is connected to $z$
neighbors at shell $l=1$. Each of the nodes in shell $l>0$ is
connected to $(z-1)$ nodes in shell $l+1$. This construction may
stop at shell $l=L$, or continue indefinitely (Fig.1).
\begin{figure}
\begin{picture}(135,135)
\includegraphics{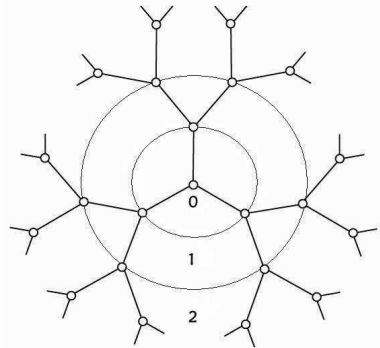}
\end{picture}
 \caption{Cayley tree with coordination number $z=3$.}
 \label{cayley}
\end{figure}
The interior of an infinite Cayley tree is called the Bethe
lattice. Due to distinctive topological structure of Cayley tree,
the reaction-diffusion processes on this graph may be integrable.

The diffusion-controlled process of cluster growth, introduced by
Witten and Sander \cite{14}, has been studied on Cayley tree in
\cite{15}, and in \cite{16}, the two-particle annihilation
reaction for immobile reactant has been studied on Bethe lattice.
Diffusion-limited coalescence, $A+A\rightarrow A$, and
annihilation, $A+A\rightarrow 0$ \cite{17}, and random sequential
adsorption \cite{18} have also been studied on Cayley tree. Also
the reaction-diffusion processes on Cayley tree, solvable through
the empty interval method, have been studied in \cite{19}.

The present paper is devoted to the study of some integrable
reaction-diffusion processes on Cayley tree. Denoting the number
of particles on $l$-th shell by $m_{l}$, we seek the situations in
which the probability $P(m_{0},m_{1},m_{2},\cdots,m_{L};t)$ can be
calculated exactly. $l_{\rm {max}}=L$ is the last shell of the
Cayley tree. By considering the most general reactions on Cayley
tree in section 2, it is shown that we must restrict ourselves to
three distinct creation reactions so that a subclass of
$L+1$-shell probabilities $P(m_{0},m_{1}\cdots,m_{L};t)$, i.e.
$P_{l}(m;t)\equiv P(0,0,\cdots,0,m_l=m,0,\cdots,0;t)$, becomes
solvable. The exact solution of $P_{l}(m;t)$ is obtained in
section 3. Finally in conclusion section it is shown that the
diffusion process can also lead to an integrable model, if one
restrict himself to one particle probabililities $P_{l}(1;t)$, and
if an extra trapping reaction exists at the central node. An ideal
spherical trap, surrounded by a swarm of Brownian particles, is a
fundamental model which had been first presented by von
Smoluchowski \cite{20}, and has been generalized to a mobile trap
in \cite{21}. Also the energy transfer process in dendrimer
supermolecule on the Cayley tree with a central trap has been
discussed in \cite{n1}. These two models are only integrable
models in this context.

It must be added that the procedure introduced in this paper can
be easily extended to the situations in which the number of
occupied shells is greater than one, but there must exist at least
one empty shell between any two of them.
\section{The integrable models}
The most general reactions of single species models with
nearest-neighbor interactions are
\begin{eqnarray}\label{1}
1: \circ\circ&\rightarrow &\bullet\circ,\ \ \ {r_{1}} \ \ \ \ \ \
\ \ \ \ \ \ \ \ 4: \bullet\circ\rightarrow\circ\bullet,\ \ \ {
r_{4}}\cr 2: \circ\circ&\rightarrow &\bullet\bullet,\ \ \ { r_{2}}
\ \ \ \ \ \ \ \ \ \ \ \ \ \ 5: \bullet\circ\rightarrow\circ\circ,\
\ \ {r_{5}}\cr
 3: \circ\bullet&\rightarrow &\bullet\bullet,\ \ \ { r_{3}} \
\ \ \ \ \ \ \ \ \ \ \ \ \ 6: \bullet\bullet\rightarrow
\circ\circ,\ \ \ { r_{6}}\cr && \ \ \ \ \ \ \ \ \ \ \ \ \ \ \ \ \
\ \ \ \ \ \ \ 7: \bullet\bullet\rightarrow\bullet\circ,\ \ \ {
r_{7}}
\end{eqnarray}
where an empty (occupied) site is denoted by $\circ(\bullet)$. The
$r_{i}$s are reaction rates and there is no distinction between
left and right. Our main goal is to find the situations which lead
to integrable models. By integrability, we mean the possibility of
the exact calculation of probabilities $P(m_{0},m_{1},\cdots;t)$.

The necessary condition for achieving this goal is that the
occupation numbers of shells (the number of occupied sites of
every shell) must be the only parameters needed to characterize
the configurations. Clearly this is not the case if two adjacent
shells are occupied. For example consider $P(0,1,2;t)$ and the
reaction (1) of eq.(1). In Figs.(2) and (3), the source terms of
two different configurations (a) and (b), with same occupation
numbers $(0,1,2)$, are shown. As it is seen, the configuration (a)
has two source terms, while the configuration (b) has three. The
number behind each configuration is its multiplicity, i.e. the
number of ways which this configuration can lead to the desired
state. This shows that if two adjacent shells are occupied, it is
not possible to write the same evolution equation for different
configurations of a given occupation numbers. Therefore, we only
consider the situations in which there is only one non-zero
occupation number, i.e. $m_{i}=m\delta_{il}$. As mentioned
earlier, it is possible to extend our procedure to the situations
in which there is at least one empty shell between any two
occupied shells.
\begin{figure}
\begin{picture}(75,75)
\includegraphics{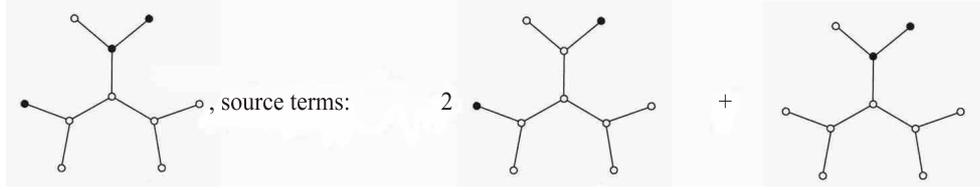}
\end{picture}
\caption{The source terms of configuration (a) of ($0,1,2$)
occupation numbers.}
\end{figure}
\begin{figure}
\begin{picture}(70,70)
\includegraphics{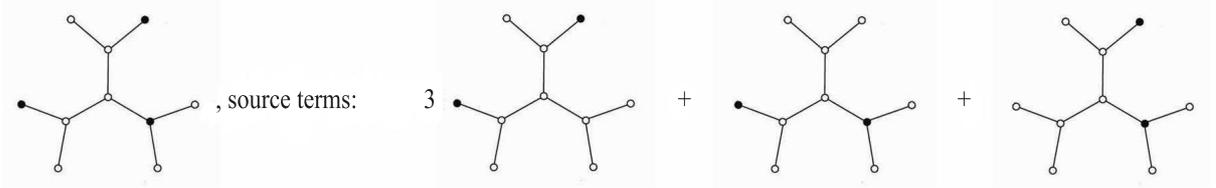}
\end{picture}
\caption{The source terms of configuration (b) of ($0,1,2$)
occupation numbers.}
\end{figure}

Considering the probabilities $P_{l}(m;t)=
P(0,0,\cdots,0,m_l=m,0,\cdots,0;t)$, we must yet check whether all
the reactions of eq.(\ref{1}) are acceptable or not. By the term
acceptable, we mean to have two properties. First, the
determination of occupation number $m$ is sufficient for having a
unique evolution equation, irrespective of particles'
distribution. Second, all the terms of evolution equation are
expressible in terms of $P_{k}(n;t)$s. Now it can be seen that
this is not the case for reactions (4)-(7) of eq.(\ref{1}). This
is because in all these cases, the source terms of $P_l(m;t)$s are
$2$-shell probabilities $P(m_{l-1},m_{l};t)$ and
$P(m_{l},m_{l+1};t)$, which can not be expressed in terms of
$P_{k}(n;t)$s. For these reactions, the master equations are:
\begin{eqnarray}\label{a}
{\partial\over{\partial t}} P_l (m;t) &=& a_4
P(m_{l-1}=1,m_l=m-1;t)+b_4 P(m_l=m-1,m_{l+1}=1;t)\cr &&
-s_4P_l(m;t)\ \ \ ( \ {\rm for}\ \ r=r_4),
\end{eqnarray}
\begin{eqnarray}\label{b}
{\partial\over{\partial t}} P_l (m;t) &=& a_5
P(m_{l-1}=1,m_l=m;t)+b_5 P(m_l=m,m_{l+1}=1;t)\cr && +c_5
P_l(m+1;t)-s_5 P_l(m;t)\ \ \ ( \ {\rm for}\ \ r=r_5),
\end{eqnarray}
\begin{eqnarray}\label{c}
{\partial\over{\partial t}} P_l (m;t) &=&a_6
P(m_{l-1}=1,m_l=m+1;t)\cr && +b_6 P(m_l=m+1,m_{l+1}=1;t)\ \ \ ( \
{\rm for}\ \ r=r_6),
\end{eqnarray}
and
\begin{eqnarray}\label{d}
{\partial\over{\partial t}} P_l(m;t) &=& a_7
P(m_{l-1}=1,m_l=m;t)\cr && +b_7 P(m_l=m,m_{l+1}=1;t)\ \ \ ( \ {\rm
for}\ \  r=r_7),
\end{eqnarray}
respectively. The parameters $a_i,b_i,c_i$ and $s_i$ are some
constants. The case $m=1$ is an exception which will be discussed
in section 4.

So the integrable model, through the $P_{l}(m;t)$ probabilities,
is a model in which the allowed reactions are reactions (1)-(3) of
eq.(\ref{1}). The evolution equation of $P_{l}(m;t)$ on a Cayley
tree with coordination number $z$ and $L+1$ shells then becomes
(for $l=0,1,2,\cdots,L-1$)
\begin{eqnarray}\label{3}
{\partial\over{\partial t}} P_{l} (m;t) &=& (n_{l}-m+1)zr_{1}P_{l}
(m-1;t)\cr &&-[2(\sum^{L}_{l'=1} n_{l'}-m z)r_{1}+(
\sum^{L}_{l'=1}n_{l'}-m z)r_{2}+mzr_{3}]P_{l}(m;t).
\end{eqnarray}
In above equation
\begin{equation}\label{4}
n_{l}=\left \{
\begin{array}{ccc}
z(z-1)^{l-1}&{\rm for}&l\geq1\\
1&{\rm for}&l=0
 \end{array}
 \right .
 \end{equation}
 is the number of sites of the $l$-th shell. The first term in the
 right-hand site of eq.(\ref{3}) is the source term of reaction (1),
 and $2(\sum^{L}_{l'=1}n_{l'}-mz)r_{1}P_{l}(m;t)$ is its sink
 term. This is because $\sum^{L}_{l'=1}n_{l'}-mz$ is the number of empty
 pairs ($\circ \ \circ$) of a Cayley tree with $m$ particles on shell $l$, which each
 of them can change to either $\bullet \ \circ$ or $\circ \ \bullet$
 with rate $r_{1}$. The reaction (2) has no source term for
 $P_{l}(m;t)$, and its sink term is clearly
 $(\sum^{L}_{l'=1}n_{l'}-mz)r_{2}P_{l}(m;t)$. Finally the reaction
 {3} does not have source term, and its sink term is
 $mzr_{3}P_{l}(m;t)$ because each of the neighbors of each
 particles of $l$-th shell can be changed via
 $\circ\bullet\rightarrow\bullet\bullet$ reaction.

 For $l=L$, since there is no upper shell, the number of neighboring sites of
 each occupied site is one, instead of $z$. Therefore for
 $P_{L}(m;t)$, the evolution equation is same as eq.(\ref{3}), when $z$ is replaced by 1.

 \section{Solutions of the master equation}
In this section we try to solve the master equation
\begin{eqnarray}\label{5}
{\partial\over{\partial t}} P_l(m;t) &=&
(n_l-m+1)zr_1P_l(m-1;t)\nonumber \\
&&-[(r_3-r_2-2r_1)m z+(2r_1+r_2)N]P_l(m;t)\ \ \ (l<L),
\end{eqnarray}
 with
 \begin{equation}
 N\equiv\sum^{L}_{l'=1}n_{l'},
 \end{equation}
 in two cases $r_{3}-r_{2}-2r_{1}\neq 0$ and
 $r_{3}-r_{2}-2r_{1}=0$. For $l=L$, we must set $z=1$ in
 eq.(\ref{5}).
\subsection{The case $r_{3}\neq r_{2}+2r_1$ }
To solve eq.(\ref{5}), we use a recursive method. We first
consider the case $m=0$, where eq.(\ref{5}) is reduced to
\begin{equation}\label{6}
{\partial\over {\partial  t}} P_{l}(0;t) =
-N(2r_{1}+r_{2})P_{l}(0;t),
\end{equation}
with solution
\begin{equation}\label{7}
P_{l}(0;t)= P_{l}(0)e^{-N(2r_{1}+r_{2})t}.
\end{equation}
$P_{l}(0)\equiv P_{l}(0;t=0)$ is the probability of finding no
particle in shell $l$  at $t=0$. Inserting eq.(\ref{7}) in $m=1$
case of eq.(\ref{5}), results in $P_{l}(1;t)$ as follows
\begin{eqnarray}\label{8}
 P_l(1;t)&=& \left[P_l(1)-\frac{n_lr_1P_l(0)}{r_3-r_2-2r_1}\right]
 e^{-[z(r_3-r_2-2r_1)+N(2r_1+r_2)]t}\nonumber \\ && +\frac{n_lr_1P_l(0)}
 {r_3-r_2-2r_1}e^{-N(2r_1+r_2)t}.
\end{eqnarray}
Continuing this method, one can deduce the following general
expression for $P_l(m;t)$:
\begin{eqnarray}\label{9}
 P_{l}(m;t)&=&\sum^{m}_{k=0}\sum^{k}_{j=0}\frac{(n_{l}-j)!}{(n_{l}-m)!(m-k)!(k-j)!}
 \left(\frac{r_{1}}{r_{3}-r_{2}-2r_{1}}\right)^{m-j}(-1)^{k-j}\nonumber \\&& \times P_{l}(j)
 e^{-[j z(r_{3}-r_{2}-2r_{1}) +N(2r_{1}+r_{2})]t}\ \ \ (l<L),
\end{eqnarray}
in which, as we will show, $P_{l}(j)\equiv P_{l}(j;t=0)$.

It can be seen that for $m=0$ and $m=1$, eq.(\ref{9}) leads to
eqs.(\ref{7}) and (\ref{8}), respectively. To prove that
eq.(\ref{9}) is the solution of eq.(\ref{5}), we insert it in
eq.(\ref{5}). The summations in the left-hand side and in the
second term of the right-hand side of eq.(\ref{5}) are both
$\sum^{m}_{k=0}\sum^{k}_{j=0}$. If we write them as
$\sum^{m}_{j=0}+\sum^{m-1}_{k=0}\sum^{k}_{j=0}$, then it can be
seen that the $\sum^{m}_{j=0}$ terms in both sides are the same,
and therefore they are cancelled. So in all three terms of
eq.(\ref{5}), the summations have the  same form
$\sum^{m-1}_{k=0}\sum^{k}_{j=0}$, which after some simple
calculations, it can be shown that the remaining terms are
cancelled, which proves the expression (\ref{9}) satisfies the
master equation (\ref{5}). Note that because of the $(n_l-m)!$
factor in the denominator of eq.(\ref{9}), $P_l(m;t)$ satisfies
\begin{equation}\label{10}
 P_l(m>n_l;t)=0,
\end{equation}
which shows the correct behavior of eq.(\ref{9}).

It is also necessary to prove the physical interpretation of
$P_l(j)$ as $P_l(j;t=0)$. To see this, we consider eq.(\ref{9}) at
$t=0$:
\begin{eqnarray}\label{11}
P_l(m;t=0) &=&
\sum^{m}_{j=0}\sum^{m}_{k=j}\left(\frac{r_1}{r_3-r_2-2r_1}\right)^{m-j}
\frac{(n_l-j)!}{(n_l-m)!}P_l(j)\cr &&
\times\frac{(-1)^{k-j}}{(m-k)! (k-j)!} ,
\end{eqnarray}
in which we use the equality:
\begin{equation}\label{12}
\sum^{m}_{k=0}\sum^{k}_{j=0} A(j,k)=\sum^{m}_{j=0}\sum^{m}_{k=j}
A(j,k).
\end{equation}
Using $k'=k-j$, eq.(\ref{11}) becomes
\begin{eqnarray}\label{13}
P_l(m;t=0) &=&
\sum^{m}_{j=0}\left(\frac{r_1}{r_3-r_2-2r_1}\right)^{m-j}
\frac{(n_l-j)!}{(n_l-m)!}P_l(j)\cr && \times
\sum^{m-j}_{k'=0}\frac{(-1)^{k'}}{k'! (m-j-k')!} .
\end{eqnarray}
From binomial expansion
\begin{equation}\label{14}
(a-b)^{n}=\sum^{n}_{k'=0}\frac{(-1)^{k'}n! }{k'!
(n-k')!}a^{n-k'}b^{k'},
\end{equation}
one finds for $a=b$
\begin{equation}\label{15}
0=n!a^{n}\sum^{n}_{k'=0}\frac{(-1)^{k'}}{k'! (n-k')!}.
\end{equation}
So the last summation of eq.(\ref{13}) is zero for all $m-j\neq
0$, or $j=0,1,\cdots,m-1$. The only remaining term is $j=m$, which
results in
\begin{equation}\label{16}
P_l(m;t=0)=P_l(m).
\end{equation}
This completes the proof of eq.(\ref{9}) as the exact solution of
the master equation (\ref{5}), with $P_l(j)$s as the initial
values of probabilities. The probability $P_L(m;t)$ is found from
(\ref{9}), by taking $l=L$ and $z=1$.

Eq.(\ref{9}) gives the probability of finding $m$ particles at
time $t$ on shell $l$, when all other shells are empty, if we
begin with any number of particles on shell $l$ at $t=0$. Of
course for a specific initial condition, i.e.
$P_l(j;t=0)=\delta_{j,j_0}$, only one term of summation
$\sum^{k}_{j=0}$ survives.

\subsection{The case $r_3=r_2+2r_1$}
For the case $r_3=r_2+2r_1$, the master equation (\ref{5}) becomes
\begin{equation}\label{17}
{\partial\over{\partial t}}
P_l(m;t)=(n_l-m+1)zr_1P_l(m-1;t)-Nr_3P_l(m;t)\ \ \ (l<L).
\end{equation}
Using the method of the previous case, the solution of above
equation is found to be
\begin{equation}\label{18}
P_l(m;t)=e^{-Nr_3t}\sum^{m}_{j=0}\frac{(n_l-j)!}{(m-j)!
(n_l-m)!}P_l(j)(r_1zt)^{m-j}\ \ \ (l<L).
\end{equation}
It can be easily shown that the solution (\ref{18}) satisfies
(\ref{17}) and has the desired properties $P_l(m>n_l;t)=0$ and
$P_l(m;t=0)=P_l(m)$. For $l=L$, the solution is found from
eq.(\ref{18}), by taking $l=L$ and $z=1$.

\section{Conclusion}
To study the integrable reaction-diffusion processes on a
more-than-one-dimensional lattices, the most general interactions
(\ref{1}) have been considered on a Cayley tree with coordination
number $z$. It has been shown that among the probability
functions, the probabilities $P_l(m;t)$s are independent of
distribution of particles on various shells, and may lead to
integrable models if we consider two situations. The first one is
the creation-reactions (1)-(3) of eq.(\ref{1}), with master
equation (\ref{3}). The $P_l(m;t)$s for two cases $r_3\neq
r_2+2r_1$ and $r_3=r_2+2r_1$ have been considered in section 3,
with the final exact results (\ref{9}) and (\ref{18}),
respectively.

As is clear from eqs.(\ref{a})-(\ref{d}), in the case $m=1$, the
master equations (\ref{b})-(\ref{d}) still contain the $2$-shell
probabilities, but for diffusion process $r=r_4$, only the
one-shell probabilities (i.e. one  point functions) appear in
eq.(\ref{a}), which may lead to an integrable model.

The master equation of $P_l(1;t)$, with $l>1$, for diffusion
process on a Cayley tree with coordination number $z$ is:
\begin{equation}\label{19}
{\partial\over{\partial t}} P_l(1;t)=
(z-1)P_{l-1}(1;t)+P_{l+1}(1;t)-zP_l(1;t),
\end{equation}
in which the diffusion rate $r_4$ has been absorbed in the
rescaling of time $t$. The evolution equations of $P_0(1;t)$ and
$P_1(1;t)$ are
\begin{equation}\label{20}
{\partial\over{\partial t}} P_0(1;t)=P_1(1;t)-zP_0(1;t),
\end{equation}
and
\begin{equation}\label{21}
{\partial\over{\partial t}} P_1(1;t)=zP_0(1;t)+P_2(1;t)-zP_1(1;t),
\end{equation}
respectively. These equations take a form similar to
eq.(\ref{19}), provided one defines
\begin{equation}\label{22}
P_{-1}(1;t):=0,
\end{equation}
\begin{equation}\label{23}
P_0(1;t):=0.
\end{equation}
The first boundary condition is trivially satisfied, since
$P_l(m;t)$s have been defined for $l\geq 0$. But the second one
indicates a trapped reaction at the origin. So eq.(\ref{19}), with
arbitrary $l$, defines an integrable model on a Cayley tree if the
particles have diffusion process, and if there exists a trap at
the root of the lattice. When the reactions are coalescence
$A+A\rightarrow A$ and annihilation $A+A\rightarrow 0$, the
density of particles in shell $l$, i.e. $\rho_l(t)$, has been
calculated in \cite{17} for the same situation, that is a trap at
the root of a Cayley tree.

It can be shown that the solution of master equation (\ref{19})
with boundary condition (\ref{23}) is
\begin{equation}\label{37}
P_l(1;t)=(z-1)^{{(l-l_0)}/{2}}\left[I_{l-l_0}(2(z-1)^{{1}/{2}}t)-I_{l+l_0}(2(z-1)^{{1}/{2}}t)\right]e^{-zt},
\end{equation}
in which $I_n$ is the n-th order modified Bessel function.

Finally it must be added that it remains one more case which may
lead to an integrable model. Looking at eqs.(\ref{a})-(\ref{d}),
it is clear that if one takes $m=0$, only the master equation
(\ref{b}) leads to an equation which only consists of one-point
functions, i.e.
\begin{equation}\label{42}
{\partial\over{\partial t}}
P_l(0;t)=a_5P_{l-1}(1;t)+b_5P_{l+1}(1;t)+c_5P_l(1;t)-s_5P_l(0;t).
\end{equation}
It can be shown that $s_5=0$. But the point is that the
determination of $P_l(0;t)$s depends on the evaluation of
$P_l(1;t)$s, which can not be calculated. This is because the
master equation of $P_l(1;t)$s contains the two-point
probabilities $P(1,1;t)$ and therefore is not closed. Therefore
the two considered models are the only cases which can be exactly
solved.

{\bf Acknowledgement:} This work was partially supported by a
grant from the "Elite National Foundation of Iran" and
a grant from the University of Tehran.\\ \\

\end{document}